\documentclass[12pt,twoside]{article}
\usepackage{cmp2e}

\usepackage{graphicx}

\newcommand{\be}{\begin{equation}}
\newcommand{\ee}{\end{equation}}
\newcommand{\bea}{\begin{eqnarray}}
\newcommand{\eea}{\end{eqnarray}}

\newcommand{\itDelta}{{\it \Delta}}

\newcommand{\itLambda}{{\Lambda}}

\newcommand{\bra}{\langle}
\newcommand{\ket}{\rangle}
\newcommand{\dbra}{\bra \! \bra}
\newcommand{\dket}{\ket \! \ket}

\newcommand{\bq}{{\bar q}}

\newcommand{\rT}{{\rm T}}

\newcommand{\rR}{{\rm R}}

\newcommand{\rRe}{{\rm Re}}

\newcommand{\rS}{{\rm S}}

\title[Analysis of accelerations in turbulence]{Analysis of accelerations 
in turbulence based on generalized statistics}
\author[T.~Arimitsu and N.~Arimitsu]{Toshihico Arimitsu\refaddr{label1},
        Naoko Arimitsu\refaddr{label2}}
\addresses{
\addr{label1} Institute of Physics, University of Tsukuba,
Ibaraki 305-8571, Japan (arimitsu@cm.ph.tsukuba.ac.jp)
\addr{label2} Graduate School of EIS, Yokohama Nat'l.\ University,
Kanagawa 240-8501, Japan (arimitsu@ynu.ac.jp)
}
\begin{document}

\maketitle

\begin{abstract}
An analytical expression of probability density function (PDF) of
accelerations in turbulence is derived with the help of the statistics
based on generalized entropy (the Tsallis entropy or the R\'{e}nyi entropy).
It is revealed that the derived PDF explains the one obtained 
by Bodenschatz et al.\ 
in the measurement of fluid particle accelerations in fully developed turbulence
at $R_\lambda = 970$.
\pacs 47.27.-i, 47.53.+n, 47.52.+j, 05.90.+m.
\end{abstract}


The multifractal analysis of turbulence by the 
statistics based on the generalized entropy of R\'{e}nyi's 
or of Tsallis' has been developed by the present authors 
\cite{AA,AA1,AA2,AA3,AA4,AA5,AA6,AA7,AA8}.
The R\'{e}nyi entropy \cite{Renyi} has the extensive character as 
the usual thermodynamical entropy does, whereas 
the Tsallis entropy \cite{Tsallis88,Tsallis99,Havrda-Charvat}
is non-extensive.
The multifractal analysis belongs to the line of study 
based on a kind of {\it ensemble} theoretical approaches that, 
starting from the log-normal model \cite{Oboukhov62,K62,Yaglom},
continues with the $\beta$-model \cite{Frisch78},
the p-model \cite{Meneveau87a,Meneveau87b},
the 3D binomial Cantor set model \cite{Hosokawa91} and so on.
After a rather preliminary investigation of the p-model \cite{AA},
we developed further to derive the analytical expression for 
the scaling exponents of velocity structure function \cite{AA1,AA2,AA3,AA4}, and to
determine the probability density function (PDF) of velocity fluctuations 
\cite{AA4,AA5,AA6,AA7} and of velocity derivative \cite{AA8} 
by a self-consistent statistical mechanical approach.

In this paper, we will derive the formula for the PDF of the accelerations
of a fluid particle in fully developed turbulence by means of 
the multifractal analysis. With the theoretical PDF, we
will analyze the PDF of accelerations at $R_\lambda = 970$ 
(the Taylor microscale Reynolds number) obtained in 
the Lagrangian measurement of particle accelerations 
that was realized by Bodenschatz and co-workers~\cite{EB01a,EB01b} 
by raising dramatically the spatial and temporal measurement resolutions
with the help of the silicon strip detectors.

We assume that the turbulent flow, satisfying the Navier-Stokes equation
\be
\partial {\vec u}/\partial t
+ ( {\vec u}\cdot {\vec \nabla} ) {\vec u} 
= - {\vec \nabla} \left(p/\rho \right)
+ \nu \nabla^2 {\vec u}
\label{N-S eq}
\ee
of an incompressible fluid,
consists of a cascade of eddies 
with different sizes $\ell_n = \delta_n \ell_0$ 
where $\delta_n = 2^{-n}$ $(n=0,1,2,\cdots)$.
The quantities $\rho$, $p$ and $\nu$ represent, respectively, the mass density, 
the pressure and the kinematic viscosity.
The acceleration $\vec {\mathrm{a}}$ of a fluid particle is given by
the substantive time derivative of the velocity:
$
{\vec {\mathrm a}} = \partial {\vec u}/\partial t
+ ( {\vec u}\cdot {\vec \nabla} ) {\vec u}
$.
At each step of the cascade, say at the $n$th step, eddies break up into 
two pieces producing an energy cascade with the energy-transfer rate
$\epsilon_n$ that represents the rate of transfer of energy per unit mass 
from eddies of size $\ell_n$ to those of size $\ell_{n+1}$
(the energy cascade model).
The Reynolds number $\rRe$ of the system is given by 
$
{\rm Re} = \delta u_0 \ell_0/\nu = ( \ell_0/\eta )^{4/3}
$
with the Kolmogorov scale~\cite{K41} 
$
\eta = ( \nu^3/\epsilon )^{1/4}
$
where $\epsilon$ ($=\epsilon_0$) is 
the energy input rate to the largest eddies with size $\ell_0$.\footnote{
The velocity fluctuation $\delta u_n$ is defined by
$
\delta u_n = \vert u(\bullet + \ell_n) - u(\bullet) \vert
$
where $u$ represents a component of the velocity field ${\vec{u}}$.
}
Introducing the {\it pressure} (divided by the mass density) difference 
$\delta p_n$ at two points separated by the distance $\ell_n$, 
i.e., 
$
\delta p_n = \vert p/\rho(\bullet + \ell_n) - p/\rho(\bullet) \vert
$,
and the acceleration 
$
\mathrm{a}_n = \delta p_n / \ell_n
$
belonging to the $n$th step in the energy cascade,
one can estimate accelerations by 
$
\vert \vec{\mathrm{a}} \vert = \lim_{n \rightarrow \infty} \mathrm{a}_n
$.
For high Reynolds number $\rRe \gg 1$, or for the situation where 
effects of the kinematic viscosity $\nu$ can be neglected compared with
those of the turbulent viscosity, the Navier-Stokes equation (\ref{N-S eq})
is invariant under 
the scale transformation~\cite{Frisch-Parisi83,Meneveau87b}:
${\vec r} \rightarrow \lambda {\vec r}$, 
${\vec u} \rightarrow \lambda^{\alpha/3} {\vec u}$, 
$t \rightarrow \lambda^{1- \alpha/3} t$ and 
$\left(p/\rho\right) \rightarrow \lambda^{2\alpha/3} \left(p/\rho\right)$.
The exponent $\alpha$ is an arbitrary real quantity which specifies the degree
of singularity in the acceleration for $\alpha < 1.5$, i.e.,
$
\lim_{n \rightarrow \infty} \mathrm{a}_n 
= \lim_{\ell_n \rightarrow 0} \delta p_n/\ell_n
\sim \lim_{\ell_n \rightarrow 0} \ell_n^{(2\alpha/3)-1}
\rightarrow \infty
$
which can be seen with the relation
$
\delta p_n / \delta p_0 = (\ell_n / \ell_0)^{2\alpha/3}
\label{p-alpha}
$.

The multifractal analysis rests on the assumption that the distribution of
the exponent $\alpha$ is multifractal, and that the probability 
$
P^{(n)}(\alpha) d\alpha
$
to find, at a point in physical space, an eddy of size $\ell_n$ having
a value of the degree of singularity in the range 
$
\alpha \sim \alpha + d \alpha
$
is given by~\cite{AA1,AA2,AA3,AA4}
\bea
P^{(n)}(\alpha) \propto \left[ 1 - (\alpha - \alpha_0)^2 \big/ (\itDelta \alpha )^2 
\right]^{n/(1-q)}
\label{Tsallis prob density}
\eea
with 
$
(\itDelta \alpha)^2 = 2X \big/ [(1-q) \ln 2 ]
$.
The range of $\alpha$ is $\alpha_{\rm min} \leq \alpha \leq \alpha_{\rm max}$ with
$\alpha_{\rm min} = \alpha_0 - \itDelta \alpha$ and 
$\alpha_{\rm max} = \alpha_0 + \itDelta \alpha$.
Here, we assume that the distribution function at the $n$th 
multifractal depth has the structure
$
P^{(n)}(\alpha) \propto [P^{(1)}(\alpha)]^n
$.
This is consistent with the relation \cite{Meneveau87b,AA4}
$
P^{(n)}(\alpha) \propto \delta_n^{1-f(\alpha)}
$
that is a manifestation of scale inveriance and 
reveals how densely each singularity, labeled by $\alpha$, 
fills physical space.
Within the present model, the multifractal spectrum $f(\alpha)$ is 
given by~\cite{AA1,AA2,AA3,AA4}
\be
f(\alpha) = 1 + (1-q)^{-1} \log_2 \left[ 1 - \left(\alpha - \alpha_0\right)^2
\big/ \left(\Delta \alpha \right)^2 \right].
\label{Tsallis f-alpha}
\ee

To make the paper self-contained, we put here its brief derivation.
The distribution function (\ref{Tsallis prob density}) is 
derived by taking an extremum of the generalized entropy,
the R\'{e}nyi entropy~\cite{Renyi} 
$
S_{q}^{\rR}[P^{(1)}(\alpha)] = \left(1-q \right)^{-1} 
\ln \int d \alpha P^{(1)}(\alpha)^{q}
\label{SqR-alpha}
$
or the Tsallis entropy~\cite{Tsallis88,Tsallis99,Havrda-Charvat}
$
S_{q}^{\rT}[P^{(1)}(\alpha)] = \left(1-q \right)^{-1}
\left(\int d\alpha \ P^{(1)}(\alpha)^{q} -1 \right)
\label{SqTHC-alpha}
$,
under the two constraints, i.e., the normalization of distribution function:
$
\int d\alpha P^{(1)}(\alpha) = \mbox{const.}
\label{cons of prob}
$
and the $q$-variance being kept constant as a known quantity:
$
\sigma_q^2 = (\int d\alpha P^{(1)}(\alpha)^{q} 
(\alpha- \alpha_0 )^2 ) / \int d\alpha P^{(1)}(\alpha)^{q}
\label{q-variance}
$.
In spite of the different characteristics of these entropies,
the distribution function giving their extremum
has the common structure (\ref{Tsallis prob density}).
The dependence of the parameters $\alpha_0$, $X$ and $q$ on 
the intermittency exponent $\mu$ is determined, 
self-consistently, with the help of the three independent equations, i.e.,
the energy conservation:
$
\left\bra \epsilon_n \right\ket = \epsilon
\label{cons of energy}
$,
the definition of the intermittency exponent $\mu$:
$
\bra \epsilon_n^2 \ket 
= \epsilon^2 \delta_n^{-\mu}
\label{def of mu}
$,
and the scaling relation:\footnote{
The scaling relation is a generalization of the one derived first in
\cite{Costa,Lyra98} to the case where the multifractal spectrum
has negative values \cite{AA}.
}
$
1/(1-q) = 1/\alpha_- - 1/\alpha_+
\label{scaling relation}
$
with $\alpha_\pm$ satisfying $f(\alpha_\pm) =0$.
The average $\bra \cdots \ket$ is taken with $P^{(n)}(\alpha)$.
For the region where the value of $\mu$ is usually observed,
i.e., $0.13 \leq \mu \leq 0.40$,
the three self-consistent equations are solved to give 
the approximate equations~\cite{AA7}:
$
\alpha_0 = 0.9989 + 0.5814 \mu
$,
$
X = - 2.848 \times 10^{-3} + 1.198 \mu
$
and
$
q = -1.507 + 20.58 \mu - 97.11 \mu^2 + 260.4 \mu^3 - 365.4 \mu^4 + 208.3 \mu^5
$.

Since there are two mechanisms in turbulent flow to rule 
its evolution, i.e., the one controlled by the kinematic viscosity 
that takes care thermal fluctuations, and the other by the turbulent viscosity
that is responsible for intermittent fluctuations related to the singularities 
in acceleration, it may be reasonable to assume that the probability 
$\Lambda^{(n)}(y_n) dy_n$ to find the scaled pressure fluctuations
$
\vert y_n \vert = \delta p_n /\delta p_0
$
in the range $y_n \sim y_n+dy_n$ has two independent origins:
\be
\Lambda^{(n)}(y_n) dy_n = \Lambda^{(n)}_{\rS}(\vert y_n \vert) dy_n
+ \Delta \Lambda^{(n)}(y_n) dy_n.
\label{def of Lambda}
\ee
The singular part $\Lambda^{(n)}_{\rS}(\vert y_n \vert)$ of the PDF stemmed from
multifractal distribution of the singularities,
and the correction part $\Delta \Lambda^{(n)}(y_n)$ 
from thermal dissipation and/or measurement error.\footnote{
Needless to say that each term in (\ref{def of Lambda}) is a multiple of two PDF's, 
i.e., the PDF for one of the two independent origins to realize and the conditional PDF
for a value $\omega_n$ in the range $\omega_n \sim \omega_n + d \omega_n$ to come out.
This is of course in a generalized sense in which the second correction term
may weaken the first singular contribution.
}
The former is derived through
$
\Lambda^{(n)}_{\rS}(\vert y_n \vert) dy_n \propto P^{(n)}(\alpha) d \alpha
$
with the transformation of the variables:
$
\vert y_n \vert = \delta_n^{2\alpha/3}
$.
The $m$th moments of the pressure fluctuations are given by 
\be
\dbra \vert y_n \vert^m \dket \equiv \int_{-\infty}^{\infty} dy_n  
\vert y_n \vert^m \Lambda^{(n)}(y_n)
= 2 \tilde{\gamma}^{(n)}_m
+ (1-2\tilde{\gamma}^{(n)}_0 ) \
a_{2m} \ \delta_n^{\zeta_{2m}}
\label{structure func m}
\ee
with
$
a_{3\bq} = \{ 2 / [\sqrt{C_{\bq}} ( 1+ \sqrt{C_{\bq}} ) ] \}^{1/2}
$,
$
{C}_{\bq}= 1 + 2 \bq^2 (1-q) X \ln 2
\label{cal D}
$
and
\be
2\tilde{\gamma}^{(n)}_m = \int_{-\infty}^{\infty} dy_n\ 
\vert y_n \vert^m \Delta \Lambda^{(n)}(y_n).
\ee
We used the normalization: $\dbra 1 \dket = 1$.
The quantity 
\be
\zeta_m = \frac{\alpha_0 m}{3} 
- \frac{2Xm^2}{9 \left(1+\sqrt{C_{m/3}} \right)}
- \frac{1}{1-q}\left[1-\log_2 \left(1+\sqrt{C_{m/3}} \right) \right] 
\label{zeta}
\ee
is the so-called scaling exponent of velocity structure function,
whose expression was derived first by the present authors \cite{AA1,AA2,AA3,AA4}.
The formula explains quite well experimental data \cite{AA1,AA2,AA3,AA4,AA5,AA7}.
Note that the formula is independent of the length $\ell_n$, and, therefore, 
independent of $n$.

\begin{figure}
\begin{center}
\includegraphics[width=7.5cm]{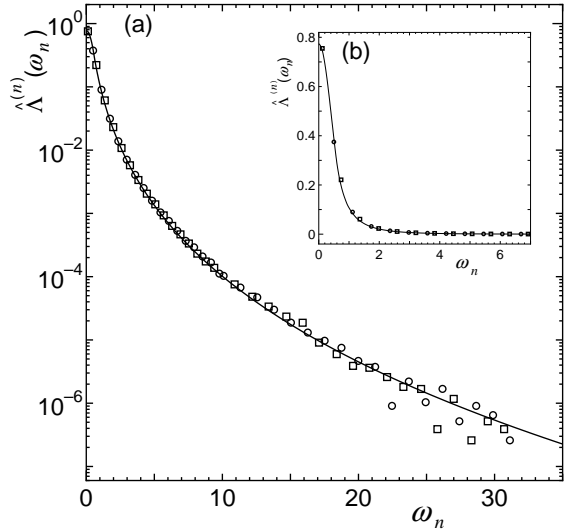}
\end{center}
\caption{\label{PDF acceleration log-linear}PDF of accelerations 
plotted on (a) log
and (b) linear scale.
Comparison between the experimentally measured PDF of 
fluid particle accelerations by 
Bodenschatz et al.\ at $R_\lambda =970$ ($\rRe = 62\ 700$) 
and the present theoretical PDF $\hat{\Lambda}^{(n)}(\omega_n)$.
Open squares are the experimental data points on the left hand side 
of the PDF, whereas open circles are those on the right hand side. 
Solid lines represent the curves given 
by the present theory (\ref{PDF accel}) with $\mu = 0.259$ ($q = 0.431$) and $n=16.4$.
}
\end{figure}

Let us derive the PDF $\hat{\Lambda}^{(n)}(\omega_n)$ of the accelerations
by dividing it into two parts with respect to $\omega_n$, i.e.,
\be
\hat{\Lambda}^{(n)}(\omega_n) = \left\{
\begin{array}{ll}
\hat{\Lambda}^{(n)}_{<\dagger}(\omega_n)
& \quad \mbox{for }\ \vert \omega_n \vert \leq \omega_n^\dagger
\\
\hat{\Lambda}^{(n)}_{\dagger<}(\omega_n)
& \quad \mbox{for }\ \omega_n^\dagger \leq \vert \omega_n \vert \leq \omega_n^{\rm max}
\end{array}
\right.
\label{PDF accel}
\ee
where the scaled variable $\omega_n$ is defined by
$
\vert \omega_n \vert = \mathrm{a}_n / \dbra \mathrm{a}_n^2 \dket^{1/2}
= \bar{\omega}_n \delta_n^{2\alpha /3 -\zeta_4 /2}
$
with
$
\bar{\omega}_n = [2 \tilde{\gamma}_2^{(n)} \delta_n^{-\zeta_4} + (1-2\tilde{\gamma}_0^{(n)} ) 
a_4 ]^{-1/2}
$, and
$
\omega_n^{\rm max} = \bar{\omega}_n \delta_n^{2\alpha_{\rm min}/3 -\zeta_4 /2}
$.
Assuming that, for smaller accelerations $\vert \omega_n \vert \le \omega_n^\dagger$, 
the contribution to the PDF comes, mainly, from
thermal fluctuations related to the kinematic viscosity or from measurement error, 
we take for the PDF 
$\hat{\Lambda}^{(n)}_{<\dagger}(\omega_n) \omega_n = 
\Lambda^{(n)}_{\rS}(y_n) dy_n
+ \Delta \Lambda^{(n)}(y_n) dy_n$ a Gaussian function, i.e.,
\be
\hat{\Lambda}^{(n)}_{<\dagger}(\omega_n) = \tilde{\Lambda}^{(n)}_{\rS}
\exp\left\{-\frac12 \left[1+\frac32 f'(\alpha^\dagger)\right] 
\left[\left(\omega_n/\omega_n^\dagger \right)^2 -1 \right]\right\}
\label{PDF less}
\ee
with
$
\tilde{\Lambda}^{(n)}_{\rS} = 3 (1-2\tilde{\gamma}^{(n)}_0)
/ (4 \bar{\omega}_n \sqrt{2\pi X \vert \ln \delta_n \vert} )
$.
On the other hand, we assume that the main contribution to 
$\hat{\Lambda}^{(n)}_{\dagger<}(\omega_n)$ may come from 
the multifractal distribution of singularities related to
the turbulent viscosity, i.e.,
$
\hat{\Lambda}^{(n)}_{\dagger<}(\omega_n) d\omega_n
= \Lambda^{(n)}_{\rS}(\vert y_n \vert) d y_n 
$:
\be
\hat{\Lambda}^{(n)}_{\dagger<}(\omega_n)
= \tilde{\Lambda}^{(n)}_{\rS} \frac{\bar{\omega}_n}{\vert \omega_n \vert}
\left[1 - \frac{1-q}{n}\ 
\frac{\left(3 \ln \vert \omega_n / \omega_{n,0} \vert\right)^2}{
8X \vert \ln \delta_n \vert} \right]^{n/(1-q)}
\label{PDF larger}
\ee
with 
$
\vert \omega_{n,0} \vert = \bar{\omega}_n \delta_n^{2\alpha_0 /3 -\zeta_4 /2}
$.
The point $\omega_n^\dagger$ was defined by 
$
\omega_n^\dagger = \bar{\omega}_n \delta_n^{2\alpha^\dagger /3 -\zeta_4 /2}
$
with $\alpha^\dagger$ being the smaller solution of 
$
\zeta_4/2 -2\alpha/3 +1 -f(\alpha) = 0
$,
at which 
$
\hat{\itLambda}^{(n)}(\omega_n^\dagger)
$
has the least $n$-dependence for large $n$.
Here, $\hat{\Lambda}^{(n)}_{<\dagger}(\omega_n)$ and 
$\hat{\Lambda}^{(n)}_{\dagger<}(\omega_n)$ 
were connected at $\omega_n^\dagger$ under the condition that they should have 
the same value and the same slope there.
The specific form of Gaussian function (\ref{PDF less}) comes out 
through the connection.

With the expression (\ref{PDF less}),
we can obtain $\Delta \Lambda^{(n)}(y_n)$, and have the formula 
to evaluate $\tilde{\gamma}_m^{(n)}$ in the form
\be
2 \tilde{\gamma}_m^{(n)} = \left(K_m^{(n)} - L_m^{(n)}\right) \Big/
\left(1 + K_0^{(n)} - L_0^{(n)}\right)
\ee
where
\bea
 K_m^{(n)} = \frac{3\ \delta_n^{2(m+1)\alpha^\dagger/3 -\zeta_4/2}}
{2\sqrt{2 \pi X \vert \ln \delta_n \vert}}
\int_0^1 dz\ z^{m} \exp\left\{-\frac12
\left[1+\frac32 f'(\alpha^\dagger)\right]\left(z^2-1\right)\right\},\\
 L_m^{(n)} = \frac{3\ \delta_n^{2m \alpha^\dagger/3}}
{2\sqrt{2 \pi X \vert \ln \delta_n \vert}}
\int_{z_{\rm min}}^1 dz\ z^{m-1} 
\left[1 - \frac{1-q}{n}\ \frac{\left(3 \ln \vert z / z_0^\dagger \vert \right)^2}{
8X \vert \ln \delta_n \vert} \right]^{n/(1-q)}
\eea
with
$
z_{\rm min} = \omega_{\rm min}/\omega_n^\dagger 
=\delta_n^{2(\alpha_{\rm max} - \alpha^\dagger)/3}
$
and
$
z_0^\dagger = \omega_{n,0}/\omega_n^\dagger 
=\delta_n^{2(\alpha_0 - \alpha^\dagger)/3}
$.
Now, the PDF of fluid particle accelerations 
(\ref{PDF accel}) is determined by the intermittency exponent $\mu$ and
the number $n$ of steps in the energy cascade which gives 
the eddy size $\ell_n$.

The comparison between the present PDF of accelerations and that 
measured in the experiment \cite{EB01a,EB01b} at $R_\lambda = 970$ 
is plotted in Fig.~\ref{PDF acceleration log-linear} on log and linear scale.
The intermittency exponent $\mu = 0.259$ and the number $n = 16.4$ of steps 
in the cascade are extracted by the method of 
least squares with respect to the logarithm of PDF's as the best fit of 
our theoretical formulae (\ref{PDF accel})
to the observed values of the PDF \cite{EB01a,EB01b} by discarding 
those points whose values are less than $\sim 2 \times 10^{-6}$ 
since they scatter largely in the log scale.
Substituting the extracted value of $\mu$ 
into the self-consistent equations, we have 
the values of parameters: $q = 0.431$, $\alpha_0 = 1.149$ and $X = 0.307$.
With these values, other quantities are determined, e.g.,
$
\itDelta \alpha = 1.248
$,
$
\alpha_{+} -\alpha_0 
= \alpha_0 - \alpha_{-} = 0.7211
$
and
$\omega_n^\dagger = 0.557$
($\alpha^\dagger = 1.005$).
We see an excellent agreement between the measured PDF of accelerations
and the analytical formula of PDF derived by 
the present self-consistent multifractal analysis.
The value of $\omega_n^\dagger$ tells us that the contribution 
from thermal fluctuation and/or measurement error is restricted to 
smaller values of $\omega_n$, i.e., less than about one half of 
its standard deviation. 
Then, the values of $\alpha$ responsible for 
the intermittency due to the scale invariance turn out to be
smaller than $\alpha^\dagger \approx 1$. 
This is the range within the condition 
$\alpha < 1.5$ in which the singularity appears in fluid particle accelerations.

\begin{table}[tbp]
\begin{center}
\caption{The values of the scaling exponents derived by 
the present theory for the acceleration
is also listed with the corresponding values for $q$, $\alpha_0$ and $X$.
}
\begin{tabular}{cc|cc|cc|cc}
$m$ & $\zeta_m$ & $m$ & $\zeta_m$ & $m$ & $\zeta_m$ & $m$ & $\zeta_m$ \\ 
  \hline
1  & 0.3661 & 6  & 1.741 & 11 & 2.599 & 16 & 3.175 \\ 
2  & 0.6989 & 7  & 1.945 & 12 & 2.731 & 17 & 3.268 \\ 
3  & 1.000  & 8  & 2.130 & 13 & 2.854 & 18 & 3.356 \\ 
4  & 1.272  & 9  & 2.299 & 14 & 2.969 & 19 & 3.438 \\
5  & 1.519  & 10 & 2.455 & 15 & 3.075 & 20 & 3.515 
  
\end{tabular}
\label{zeta m}
\end{center}
\end{table}

The values of the scaling exponents $\zeta_m$, given by (\ref{zeta}), for 
$m=1 \cdots 20$ are listed in Table~\ref{zeta m}
for future convenience in comparison with experiments or other theories.
Note that $\mu$ is related to 
$\zeta_6$ by the relation $\mu = 2 - \zeta_6$ within the present analysis. 
The flatness 
$
F_{\mathrm{a}}^{(n)} \equiv \dbra \mathrm{a}_n^4 \dket
/\dbra \mathrm{a}_n^2 \dket^2
=\dbra \omega_n^4 \dket 
$
of the PDF of accelerations has the value
$
F_{\mathrm{a}}^{(n)} = 57.8
$
which is compatible with the value of the flatness $\sim 60$ 
reported in \cite{EB01a,EB01b} as it should be.
It is quite attractive to see that the distance $r = \ell_n$ corresponding to
the extracted value $n=16.4$ reduces to $r= 0.821 \mu$m ($r/\eta = 0.0456$), 
which is close to the value of the spatial resolution $0.5 \mu$m 
($1/40$ of the Kolmogorov distance) of the measurement in \cite{EB01a,EB01b}.

We derived, in this letter, the formula of the PDF of accelerations 
(\ref{PDF accel}) within the approach of
the multifractal analysis constructed by the present authors, and
analyzed successfully the beautiful experiment
conducted by Bodenschatz and co-workers \cite{EB01a,EB01b} in the Lagrangian frame.
We expect that other data for the same system, in addition to 
the PDF of accelerations,
will be provided in the near future,
such as the intermittency exponent $\mu$,
the scaling exponents $\zeta_m$, the PDF's of the velocity fluctuations 
or of the velocity derivatives. Then, we can cross-check the validity 
of the present analysis based on the formula (\ref{PDF accel}).\footnote{
We have checked, with the help of the DNS data
reported by Gotoh et al.\ in \cite{Gotoh02},
the accuracy of the value of the intermittency
exponent extracted out of the measured PDF's.
With the formula (\ref{zeta}), we determined in \cite{AA7} 
the value $\mu = 0.240$ for the longitudinal velocity fluctuations
by fitting, with the method of least squares, 
the ten data of the scaling exponents $\zeta_m$ 
($m = 1,2,\cdots, 10$) at $\rR_\lambda = 381$.
On the other hand, we have extracted the value $\mu = 0.237$
by the method of least squares as the best fit of the theoretical
formula for the PDF of velocity derivatives, derived in \cite{AA8} 
by the multifractal analysis, to the corresponding PDF data.
}
Note that the formula for the PDF of accelerations is different from 
the one for the PDF's of velocity fluctuations or of velocity derivatives.
The empirical PDF,
$
\hat{\Lambda}_{\mathrm{emp}}(\omega) = C\exp\left\{-\omega^2/\left[\left(
1+\vert \omega \beta /\sigma \vert^\gamma \right) \sigma^2 \right]
\right\}
\label{empirical PDF}
$
with 
$\beta = 0.539$, $\gamma = 1.588$, $\sigma = 0.508$ and
$C = 0.786$
proposed in \cite{EB01a,EB01b} for the data at $R_\lambda = 970$, 
gives a line very close to the one provided by the present PDF 
$\hat{\Lambda}^{(n)}(\omega)$ of (\ref{PDF accel}) 
for the region $\vert \omega \vert < 30$ where the data of PDF exist.
They deviates, however, for $\vert \omega \vert > 30$, i.e., 
$\hat{\Lambda}_{\mathrm{emp}}(\omega) < \hat{\Lambda}^{(n)}(\omega)$. 
The extraction of the PDF of accelerations out of the DNS data 
obtained by Gotoh et al.\ \cite{Gotoh02} is one of the attractive 
investigations in order to check the validity of the present 
formula (\ref{PDF accel}) derived by the unified approach 
providing various PDF's \cite{AA4,AA5,AA6,AA7,AA8},
since the accuracy of their DNS data for PDF's is very high 
up to the order of 10$^{-10}$ \cite{AA10}. 


The authors would like to thank Prof.~C.~Tsallis for his fruitful comments 
with encouragement.
The authors are grateful to Prof.~E.~Bodenschatz for his kindness to show 
his data prior to publication.

%
%

\begin{thebibliography}{99}
\bibitem{AA} Arimitsu T., Arimitsu N. // Phys. Rev. E, 2000, vol. 61, 
         No. 3, p. 3237-3240.
\bibitem{AA1} Arimitsu T., Arimitsu N. // J. Phys. A: Math. Gen., 2000, 
        vol. 33, p. L235-L241[{\footnotesize CORRIGENDUM}: 2001, vol. 34,
         p. 673-674.
\bibitem{AA2} Arimitsu T., Arimitsu N. // Chaos, Solitons and Fractals, 
		2002, vol. 13, p. 479-489.
\bibitem{AA3} Arimitsu T., Arimitsu N. // Prog.~Theor.~Phys., 2001, vol. 105, 
		 No. 2, p. 355-360.
\bibitem{AA4} Arimitsu T., Arimitsu N. // Physica A, 2001, vol. 295, 
         p. 177-194.
\bibitem{AA5} Arimitsu N., Arimitsu T. // J. Korean Phys. Soc., 2002, vol. 40, 
         No. 6, p. 1032-1036.
\bibitem{AA6} Arimitsu T., Arimitsu N. // Physica A, 2002, vol. 305, 
         p. 218-226.
\bibitem{AA7} Arimitsu T., Arimitsu N. // J. Phys.: Condens. Matter, 2002,
		vol. 14, p. 2237-2246.
\bibitem{AA8} Arimitsu N., Arimitsu T. // Europhys. Lett., 2002, vol. 60, 
        No. 1, p. 60-65.
\bibitem{Renyi} R\'{e}nyi A. On measures of entropy and information.
		In: Proc.\ 4th Berkeley Symp.\ Maths.\ Stat.\ Prob. Vol.~1, University
		of California Press, Berkeley, Los Angeles, 1961, p. 547.
\bibitem{Tsallis88} Tsallis C. // J. Stat. Phys., 1988, vol. 52, p. 479-487.
\bibitem{Tsallis99} Tsallis C. // Braz. J. Phys., 1999, vol. 29, No.1, 
        p. 1-35;  On the related recent progresses see at http:
        //tsallis.cat.cbpf.br/biblio.htm.
\bibitem{Havrda-Charvat} Havrda J.H., Charvat F. //	Kybernatica, 1967, vol.3, 
        p. 30-35.
\bibitem{Oboukhov62} Oboukhov A.M. // J. Fluid Mech., 1962,vol. 13, p.77-81.
\bibitem{K62} Kolmogorov A.N. // J. Fluid Mech., 1962, vol. 13, p. 82-85.
\bibitem{Yaglom} Yaglom A.M. // Sov. Phys. Dokl., 1966, vol. 11, p. 26-29.
\bibitem{Frisch78} Frisch U., Sulem P-L., Nelkin M. // J. Fluid Mech., 1978, 
         vol. 87, p. 719-736.
\bibitem{Meneveau87a} Meneveau C., Sreenivasan K.R. // Phys. Rev. Lett., 1987, 
         vol. 59, p. 1424-1427.
\bibitem{Meneveau87b} Meneveau C., Sreenivasan K.R. // Nucl. Phys. 
        (Proc. Suppl.) B, 1987, vol. 2, p. 49-76.
\bibitem{Hosokawa91} Hosokawa I. // Phys. Rev. Lett., 1991, vol. 66, 
        p. 1054-1057.
\bibitem{EB01a}  Porta A.La, Voth G.A., Crawford A.M., Alexander J., 
		Bodenschatz E. // Nature, 2001, vol. 409, p. 1017-1019.
\bibitem{EB01b} Voth G.A.,  Porta A.La, Crawford A.M., Alexander J.,
		Bodenschatz E. // 2001, preprint.
\bibitem{K41} Kolmogorov A.N. // C.R.~Acad. Sci. USSR, 1941,vol. 30, 
        p. 301-305; p. 538-540.
\bibitem{Frisch-Parisi83} Frisch U., Parisi G. Turbulence and Predictability
        in Geophysical Fluid Dynamics and Climate Dynamics. 
        New York, North-Holland, 1985, p. 84.
\bibitem{Costa} Costa U.M.S., Lyra M.L., Plastino A.R., Tsallis C. 
        // Phys. Rev. E, 1997, vol. 56, p. 245-250.
\bibitem{Lyra98} Lyra M.L., Tsallis C. // Phys. Rev. Lett., 1998, vol.80, 
        p. 53-56.
\bibitem{Gotoh02} Gotoh T., Fukayama D., Nakano T. // Phys. Fluids, 2002, 
        in press.
\bibitem{AA10} Arimitsu T., Arimitsu N. // 2002, in preparation to submit.

\end{thebibliography}
\end{document}